\documentclass[twocolumn,showpacs,preprintnumbers,amsmath,amssymb]{revtex4}

\usepackage{graphicx}
\usepackage{dcolumn}
\usepackage{bm}

\begin{document}

\title{Opinion dynamics on directed small-world networks}
\author{Luo-Luo Jiang$^1$}
\author{Da-Yin Hua$^2$}
\author{Jun-Fang Zhu$^1$}
\author{Bing-Hong Wang$^1$}
\email{bhwang@ustc.edu.cn}
\author{Tao Zhou$^1$$^,$$^3$}

\affiliation{%
$^1$Department of Modern Physics, University of Science and
Technology of China, Hefei 230026, PR China\\
$^2$Department of Physics, Ningbo University, Ningbo Zhejiang
315211, PR China\\
$^3$Department of Physics, University of Fribourg, Chemin du Muse
3, CH-1700 Fribourg, Switzerland
}%

\date{\today}

\begin{abstract}
In this paper, we investigate the self-affirmation effect on
formation of public opinion in a directed small-world social
network. The system presents a non-equilibrium phase transition
from a consensus state to a disordered state with coexistence of
opinions. The dynamical behaviors are very sensitive to the
density of long-range interactions and the strength of
self-affirmation. When the long-range interactions are sparse and
individual generally does not insist on his/her opinion, the
system will display a continuous phase transition, in the opposite
case with high self-affirmation strength and dense long-range
interactions, the system does not display a phase transition.
Between those two extreme cases, the system undergoes a
discontinuous phase transition.
\end{abstract}


\maketitle

\section{Introduction}
Recently, much effort has been devoted to the studies of opinion
dynamics \cite{rew0,rew1,rew2}. The Sznajd model \cite{1int,2int},
Galam's majority rule \cite{3int,4int}, and the Axelrod
multicultural model \cite{5int} describe opinion dynamics as
individuals follow their nearest neighbors' opinion. However, the
real-life system often seems a black box to us: the outcome can be
observed, but the hidden mechanism may be not visible. If we see
many individuals hold the same opinion, we say \emph{The Spiral of
Silence} phenomenon \cite{6int} occurs. It is common in real world
that people adhere to their own opinion even opposite to most of
their friends \cite{6aint,6bint,6cint}, which we call
\emph{self-affirmation} of individuals. Traditional opinion models
fail to account for the self-affirmation effect of individuals. In
our early work, the influence of inflexible units has been
investigated in a simple social model \cite{7int}. It is found
that this kind of effect can lead to a nontrivial phase diagram.

In the real world, interactions between individuals are not only
short ranged, but also long ranged \cite{8int,9int}. The
interaction usually takes on directed feature, which means an
individual who receives influence from a provider may not effect
the provider. We use directed links to represent this kind of
relation between individuals in the directed small-world networks
proposed by S\'{a}nchez \emph{et al} \cite{10int}. In this paper,
we present an opinion dynamics model including individual
self-affirmation psychological feature and directed long-range
correlation between individuals. The parameter space can be
roughly divided into three regions, in which, respectively, we
observed continuous phase transition, discontinuous phase
transition and no phase transition.

\section{model}
In this section, we introduce a directed small-world network model
and an opinion dynamics model. We start with a two-dimensional
regular lattice, in which every node is connected with adjacent
four nodes inwardly and outwardly respectively, then, with
probability $p$, rewire each link connected outwardly with a
neighbor node to a randomly chosen nonadjacent node. In this way,
as shown in Fig. 1, a directed network with a density $p$ of
long-range links is obtained. In the network, nodes represent
individuals in the social system and the outward links represent
the influence from others. Each node connects with four nodes
outwardly which are called mates of the node. Long-range links are
used to describe the long-range correlations between individuals.

\begin{figure*}[!tp]

\scalebox{0.7}[0.5]{\includegraphics{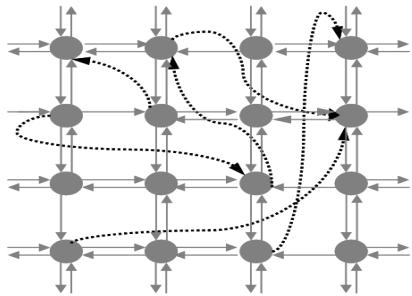}}
 \caption{Illustration of the structure of a directed small-world network for $p=0.1$ \cite{10int}.}

\end{figure*}

In the network, state of each node represents the viewpoint of the
corresponding individual, which evolves according to social
process, determined not only by other correlative surrounding
effects but also by its own character. To illustrate these cases
simply, it is supposed that there are two kinds of possible
opinions in the system, just as the agreement and disagreement in
the election, and each individual takes only one of them.
Therefore, the state of a node $i$ can be described as $\sigma_i$,
$\sigma_i\in\{+1,-1\}$. We describe the difference of $\sigma_i$
between its mates' state by the function
$W(\sigma_{{i}})=2\,\sigma_{{i}}\sum_{j=1}^{4}\sigma_{{ j}}$, in
which $\sigma_j$ $(j=1,2,3,4)$ are states of mates of node $i$. In
addition, $q$ $(0<q\leq1)$ is used to describe the probability,
with which individuals follow their mates' dominant opinion.
Meanwhile $1-q$ represents the self-affirmation probability of
individuals, with which an individual insists on his/her own
opinion though it is opposite to most of his/her mates.

According to the illumination above, we introduce the dynamic rule
as follows: $W(\sigma_{{i}})>0$ indicates that $\sigma_i$ is the
same as the majority of $\sigma_j$ $(j=1,2,3,4)$ and $\sigma_i$
overturns with probability exp$[-W(\sigma_{{i}})/T]$ which depends
on a temperature-like parameter $T$. $W(\sigma_{{i}})<0$
represents that $\sigma_i$ is opposite to the majority of
$\sigma_j$ $(j=1,2,3,4)$, and $\sigma_i$ overturns with
probability $q$. When $W(\sigma_{i})=0$, we consider that the
state of node $i$ overturns with probability $q$ also. So that the
overturning probability $P (\sigma_{i})$ of $\sigma_{i}$ is given
by
\begin{equation}
    P{(\sigma_{i})}=\left\{
    \begin{array}{cc}
     $exp[$-W(\sigma_{i})/T]$$      &\mbox{ ,  for $W(\sigma_{i})> 0$}\vspace{1mm}\\
     \emph{$q$}                          &\mbox{ ,  for $W(\sigma_{i})\leq
     0$}\vspace{1mm}.
    \end{array}
    \right.
\end{equation}
From the dynamical rule (1), we can see that, when $q=1$, the
current model restores to the network-based Ising model
\cite{10int}. However, our model is non-equilibrium because the
overturning probability of a state does not satisfy the detailed
equilibrium condition.

\section{Simulations}
In order to describe the evolution process of the model, we employ
a magnetization like order parameter

\begin{equation}
m=\mid\frac{1}
 {L^{2}}\sum_{i=1}^{{L}^{2}}\sigma_{{i}}\mid    ,
 \sigma_{i}\in\{+1,-1\}.
\end{equation}
The network size is $L\times L$ and $m$ represents the absolute
average value of the states of all nodes. An extensive Monte Carlo
numerical simulation has been performed on our model with a random
initial configuration and a periodic boundary. Results are
calculated after the system reaches a non-equilibrium stationary
state. In order to reduce the occasional errors, for network size
$L=16$, $32$, $64$, and $100$, we have averaged the result over
40000, 10000, 2000, and 1000 runs, respectively, with different
network structures under different random initial configuration.
Obviously, when $\langle m \rangle$ tends to $1$, the system
enters into an ordered state, i.e. individuals in the system reach
the opinion consensus. Meanwhile, if the system stays in a
disordered state, the order parameter scales as $\langle m \rangle
\sim \frac{1}{L}$. As shown in Fig. 2, the system reaches an
ordered state when $T$ is less than a critical temperature $T_c$.
The system displays a continuous phase transition for $p=0.1$ and
$q=0.9$, while a discontinuous phase transition for $p=0.9$ and
$q=0.9$. From the probability density functions (PDFs) of the
order parameter near the phase transition point of the phase
diagram $(p, T)$, one can distinguish between the continuous phase
transition and discontinuous phase transition clearly. According
to PDFs inserted in the upper and lower of Fig. 2, It is found
that the most probable values of $m$, which correspond to the
highest peaks of PDFs, jump little from nonzero to zero in
continuous phase transitions from Fig. 2a to Fig. 2b, while
largely in discontinuous ones from Fig. 2c to Fig. 2d. It seems
that the long-range correlations can change the nature of phase
transition.

\begin{figure*}[!tp]

\scalebox{1.3}[1.1]{\includegraphics{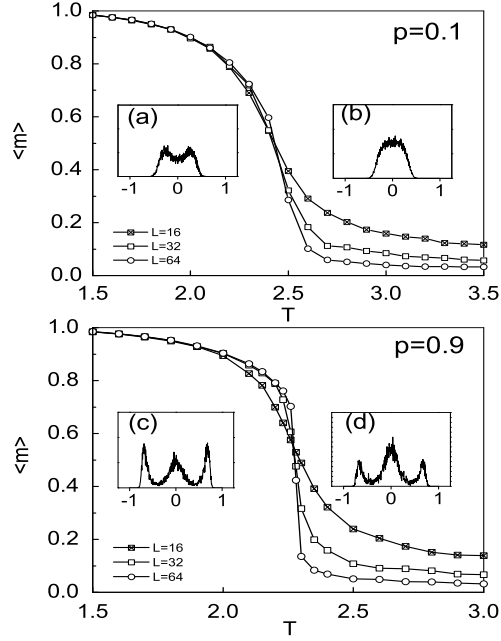}} \caption{$\langle m
\rangle$ varies with $T$ for different system sizes. The upper and
lower plots are for $p=0.1$ and $p=0.9$, with $q=0.9$ fixed.
Insets are PDFs nearby the phase transition point: (a)
$T\rightarrow T_c^-$, (b) $T\rightarrow T_c^+$, (c) $T\rightarrow
T_c^-$, (d) $T\rightarrow T_c^+$.}

\end{figure*}

Evidently, the system varies from the continuous phase transition
to discontinuous phase transition when the density of directed
long-range connections is high enough for $q=0.9$. It is natural
to ask how these topology structures influence the opinion
dynamics. To solve this problem, we define the domain size $s$ as
the number of neighborhood nodes in the same state. As showed in
Fig. 3a, it is found that the domain size $s$ distributes in a
power law, $g(s)$$\sim$$s^{-\tau}$ for $s\ll L^2$ at the critical
point, where $g(s)$ is the probability function. One can find that
there is a local maximum probability of large domain for $p=0.1$
compared to the size distribution for $p=0.9$ in Fig. 3a. Smaller
$p$ indicates more localized interactions between individuals, and
a large domain emerges more easily. Besides, we calculate the
number of time steps, $t$, during which an individual holds the
same opinion. As shown in Fig. 3b, one can find that individuals
change their own opinion for $p=0.9$ more frequently than for
$p=0.1$ at $T=0.1$, and the probability of $t$ obeys a power-law
distribution $f(t)$$\sim$$t^{-\gamma}$ $(t<t_0)$ for $p=0.1$.
Clearly, $p$ plays the key role in determining the communication
strength between different opinion domains, and individuals change
their own opinions more frequently because of the effect of
long-range connections between different opinion domains.

\begin{figure*}[!tp]
\scalebox{1.3}[1.1]{\includegraphics{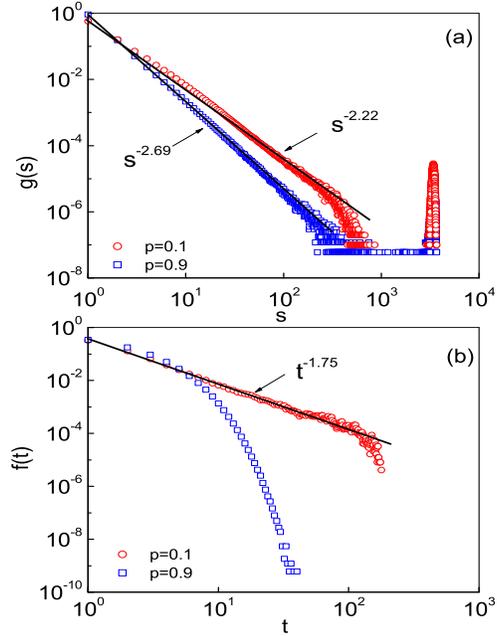}} \caption{ (Color
online.) The distribution of domain size $g(s)$ (a) and opinion
holding time $f(t)$ (b) in different networks with $q=0.9$ fixed,
(a) for $T=T_{c}$ and (b) for $T=0.1$. The data points are
obtained from $10^5$ samples with fixed network size, $L=64$.}
\end{figure*}

Fig. 4a, 4b, and 4c show the phase diagram of opinion dynamics
determined by the network structure parameter $p$ as well as the
individual self-affirmation psychology characteristic parameter
$1-q$. In Fig. 4a for q=0.9, the system displays continuous phase
transition for $p<p_c$, while discontinuous phase transition for
$p\geq p_c$. The system displays discontinuous phase transition
for $q=0.5$ in Fig. 4b. The system displays discontinuous phase
transition for $p< p_0$ and $q=0.3$ in Fig. 4c, while the system
does not have a phase transition for $p> p_0$ and $q=0.3$ in Fig.
4c. As shown in Fig. 4d, the continuous phase transition takes
place in the area $I$, the discontinuous phase transition appears
in the area $II$ and the system stays in disordered state without
phase transition in the area $III$. When both the parameters $p$
and $1-q$ are large enough, indicating weak interaction between
individuals in both local and global levels, the system keeps in
disordered completely at any temperature, i.e. the phase
transition can not take place in the system, as in the area $III$
of Fig. 4d.

\begin{figure*}

\scalebox{0.8}[0.75]{\includegraphics{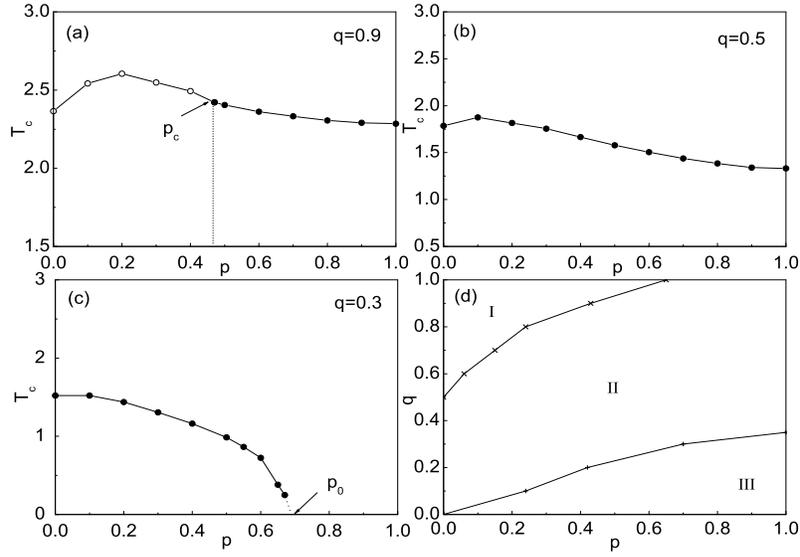}} \caption{ The phase
diagram of the model in the $p-q$ plane. Points are numerical
determinations of the critical temperatures $T_c$ for different
degrees of topological disorder $p$. The transition is continuous
for open circles points, while discontinuous for filled circles
points. Plot reports the full picture of phase diagram: the system
displays continuous phase transition in region $I$, discontinuous
phase transition in region $II$, and no phase transition in region
$III$.}

\end{figure*}

A finite-size scaling analysis is employed to study the critical
behavior of continuous phase transition for $p=0.1$ and $q=0.9$.
In the neighborhood of the critical point $T_c$, $\langle m
\rangle \propto(T_c-T)^\beta$, $(T<T_c)$, where $\beta$ is the
order parameter exponent. Besides, when $T$ is near to critical
point $T_c$ of the second order phase transition, a character
length scale $\xi$ denotes the correlation length in space.
$\xi\propto(T_c-T)^{-\nu}$, $(T<T_c)$, where $\nu$ is a
correlation length exponent in the space direction. At critical
point, various ensemble-averaged quantities depend on the ratio of
system size and the correlation length $L/\xi$. Therefore, the
order parameter $\langle m \rangle$ satisfies the scaling law in
the neighborhood of the critical point: $\langle m \rangle \propto
L^{-\beta/\nu}f[(T_c-T)L^{1/\nu}]$. At $T_c$, $\langle m \rangle
\propto L^{-\beta/\nu}$, and we obtain $\beta/\nu=0.530(5)$ for
$p=0.1$ and $q=0.9$ in Fig. 5(a). Fig. 5(b) reports $\langle m
\rangle L^{\beta/\nu}$ versus $(1-T/T_c)L^{1/\nu}$ on a
double-logarithmic plot for $q=0.1$ and q=0.9. It is shown that
with the choices $\beta/\nu=0.530 (5)$ and $\nu=0.92(1)$ the data
for different network sizes are well collapsed on a single master
curve \cite{sim2}. The slope of the line is $\beta=0.488\pm0.005$,
which gives the asymptotic behavior for $ \langle m \rangle
L^{\beta/\nu}$ as $L\rightarrow\infty$. So that, we have
$\beta=0.488(5)$, $\nu=0.92(1)$.

\begin{figure*}

\scalebox{1.3}[1.1]{\includegraphics{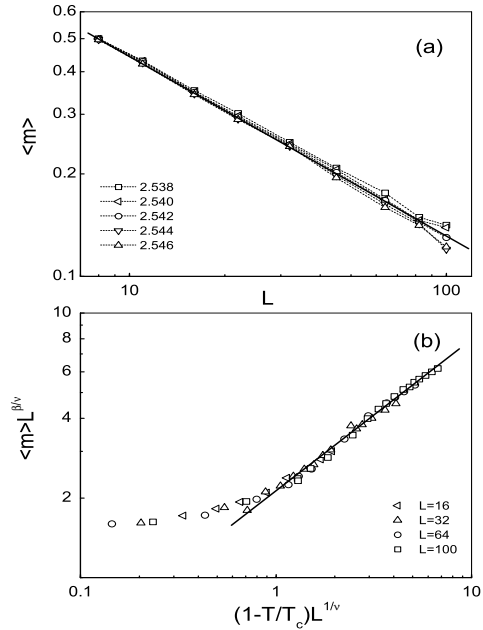}} \caption{Finite size
scaling of continuous phase transition for $p=0.1$ and $q=0.9$.
(a) A log-log plot of the order parameter $\langle m \rangle$
against $L$. (b) Double logarithmic plot of $\langle m \rangle$
$L^{\beta/\nu}$ versus $(1-T/T_c)L^{1/\upsilon}$ for $L=16$, 32,
64, and 100.}

\end{figure*}

\section{Conclusion}
In conclusion, the effect of directed long-range links between
individuals on the opinion formation is systematically explored.
The results show that the system takes on a non-equilibrium phase
transition from a consensus state to a state of coexistence of
different opinions. With increasing density of long-range links, a
continuous phase transition changes into a discontinuous one. The
reason why the phase transition behavior varies is that the
long-range links make individuals change their own opinion more
frequently. It is worth mentioning that the system keeps in a
disordered state when there are sufficient long-range links. Those
long-range interactions break the possibly local order, thus
hinder the global consensus.

In opinion dynamics, the self-affirmation psychology character
sometime may lead to polarized decision \cite{con1,con1a}.
Moreover, interaction between individuals in social system depends
on the topology of social networks \cite{con2,con3}. In
macroscopic level, the opinion dynamics is highly affected by
social structure, while in the microscopic, it is sensitive to the
dynamical mechanism of individual. Our work shows a systematic
picture of opinion dynamics, and provides a deep insight into
effects of these two factors.

\begin{acknowledgments}
The authors wish to thank Dr. Ming Zhao and Dr. Jian-Guo Liu for
their assistances in preparing this manuscript. This work is
supported by the National Basic Research Program of China (973
Program No. 2006CB705500), the National Natural Science Foundation
of China (Grant Nos. 60744003, 10575055, 10635040, 10532060 and
10472116), and the Specialized Research Fund for the Doctoral
Program of Higher Education of China.
\end{acknowledgments}

\end{document}